\documentclass[12pt,preprint]{aastex}

\shorttitle{W49A}

\shortauthors{Conti \&  Blum }

\def\aple{$\mathrel{\hbox{\rlap{\hbox{\lower4pt\hbox{$\sim$}}}\hbox{$<$}}
}$}

\def\apge{$\mathrel{\hbox{\rlap{\hbox{\lower4pt\hbox{$\sim$}}}\hbox{$>$}}
}$}

\begin{document}

\title{Near Infrared Observations of the Giant HII Region W49A: A 
Starbirth 
Cluster}

\author{Peter S. Conti}
\affil{JILA and APS Department, University of Colorado, Boulder CO 80309
\\ pconti@jila.colorado.edu}

\author{Robert D. Blum\altaffilmark{1}}
\affil{Cerro Tololo Interamerican Observatory,
Casilla 603, La Serena, Chile \\ rblum@noao.edu}

\altaffiltext{1}{Visiting Astronomer, Cerro Tololo Interamerican
Observatory, National Optical Astronomy Observatories, which is
operated by Associated Universities for Research in Astronomy, Inc.,
under cooperative agreement with the National Science Foundation}

\begin{abstract}

W49A is one of the most luminous giant H~II (GH~II) regions in our Galaxy.
This star forming complex contains numerous compact and ultra-compact (UC)
H~II regions, extending over an area of $\approx$ 15 pc.  It emits about
$10^{51}$ Lyman continuum photons per second, equivalent to the presence
of about 100 O stars, but it is completely obscured in optical wavelengths
by intervening interstellar dust.  The center holds a ``cluster'' of about
30 O stars, each within an individual UCHII region emitting free--free
emission at cm wavelengths.  Our deep $K-$band ($2.2~\mu$m) image toward
the W49A cluster reveals just two of the individual exciting stars, each
associated with a point--like radio source, but the rest are invisible.  
These O stars are so recently born as to not yet have emerged from their
natal dust cocoons, in contrast to other Galactic clusters embedded in
GH~II regions in which many of the individual massive stars are already
revealed. Plausibility arguments are made which suggest that a stellar
disc might be common during the entire UCH~II phase of massive star birth,
as it persists after accretion ceases in some stars.  Nebular emission
(e.g., from Br$\gamma$) is visible around the periphery of the central
region of W49A, along with candidate exciting stars. Star formation there
may have preceeded that in the center, or its lower density environment
may have speeded up the dispersal of the natal dust cocoons.  The W49A
cluster can serve as a template for the more luminous buried star clusters
now being found in normal galaxies and starbursts.

\end{abstract}

\section{INTRODUCTION}

W49 is a strong radio continuum source initially discovered by
\citet{We58} in his 22~cm survey. It was subsequently found to be resolved
into two components, a thermal source, W49A, to the west and a non-thermal
source, W49B, $12.5'$ to the east \citep{Met67}. The latter is a SN
remnant.  W49A ($=$ G43.2+0.0) is a giant H~II (GH~II) region and one of
the most luminous in our Galaxy (e.g., Smith et al. 1978). Given its
location in the plane of the Milky Way and distance of $11.4\pm1.2$ kpc
\citep{Get92}, W49A is obscured optically by intervening (and local)
interstellar dust.  It has been the subject of numerous investigations in
radio and infra-red (IR) wavelengths, which we shall now briefly review. 

\citet{Met67} found W49A to have two thermal components, one, A2, made up
of of small size (\aple 1 pc) and high density ($n_e$ \apge $10^4$
cm$^{-3}$) components, embedded in another, A1, of larger diameter (\apge
14 pc) and low density ($N_e = 234~$cm$^{-3}$). In a remarkably prescient
suggestion, they pointed out that the high density (A2) sources might
surround O stars still embedded in their ionized dust cocoons as had been
predicted on general principles by \citet{DH67}. Subsequent studies at
multiple radio wavelengths and higher spatial resolution by \citet{SM69};
\citet{Aet78}; \citet{Det84}; \citet{Deet97}; \citet{Deet00} and others
have identified individual point-like and spatially extended sources
(H~II regions) within W49A.  The former are interpreted as
ultra--compact~H~II (UCH~II) regions (e.g., Wood \& Churchwell 1989a),
some of which are found in a ring--like structure \citep{Wet87} that may
be rotating. This cluster of UCH~II regions is within a larger radio
structure labeled W49N; a separate cometary UCH~II source, W49S, is $2.5'$
distant and not considered further here.  The high density (A2) sources
proposed by Mezger et al. may be identified with the UCH~II regions and
their A1 source with W49N itself. 

The labeling of the individual sources follows \citet{Det84} and
\citet{Deet97}. There are roughly 30 radio point sources in the central
W49A region corresponding to about that number of OB stars. In the entire
W49A region there are about $10^{51}$ Lyc photons emitted per sec (from
Smith et al. 1978, modified to the new distance of Gwinn et al. 1992), or
about 100 O star ``equivalents'' (O7V*: Vacca 1994). Thus roughly one-third
of the excitation of the W49A region is accounted for by the UCH~II
sources in the central cluster and two-thirds must come from elsewhere
(including W49S). Associated with these hot stars must be thousands of
lower mass objects (not yet detected or possibly not yet formed). 

Most GH~II regions in our Galaxy are, like W49A, optically obscured by
dust in the Milky Way (a notable exception is NGC~3603; e.g, Drissen et 
al. 1995).  Near IR observations (e.g., in the $K-$band at
2.2~$\mu$) have the potential to penetrate dust (since $A_K$ is
$\approx$~0.1 $A_V$), enabling one to identify and classify the O type
stars exciting the H~II regions following \citet{Het96,Het97}. 
\citet{Bet99,Bet00,Bet01} have begun a program to study GH~II regions in
the near IR, using $JHK$ imaging photometry to select the candidate
exciting stars and $K-$band spectroscopy to classify them.  Some twenty
bright Galactic sources from the H~II and GH~II compilation of
\citet{Set78} were initially selected, and $K-$band photometry has been
obtained for eight of them (four others have data from the literature). 
All of these sources, aside from W49A, show the presence of a stellar
cluster in the $K-$band at the radio source position. In W49A, there is no
central ($K-$band) cluster; the ``ring'' of UCH~II regions observed in the
radio assumes that role. 

The purpose of this article is to discuss the status of the newly born
massive stars within the cluster/association represented by W49A. A brief
summary of the properties of UCH~II regions is given in \S 2. We present
new $K-$band imaging of W49A and contrast it with existing high spatial
resolution radio maps of the region in \S3.  Emission nebulae and their
possible exciting stars are visible in the $K-$band image in the outer
fringes of W49A but aside from two point sources, the stars of the central
cluster of UCH~II regions are not. An interpretation of W49A as a newly
born cluster in which the stars are still embedded in their birth cocoons
but in different stages of emergence is given in \S4. A connection to the
recently discovered ultra-dense~H~II (UDH~II) regions in 
M~33 and other galaxies
\citep{Jet01} is provided in \S5 and a summary of cluster starbirth
phenomena follows in \S6. 

\section{UCH~II REGIONS}

\subsection{General Properties}

We will first briefly review the properties of UCH~II regions, mostly
following \citet{Ch99}. The candidate objects are typically initially
identified from radio wavelength observations as thermal free--free
sources. The radio radiation comes from a very compact (\aple~0.1 pc), but
dense ($n_e$ \apge $10^4$ cm$^{-3}$), H~II region surrounding a newly born
OB star. UCH~II regions all radiate strongly in IR wavelengths from their
extensive heated dust cocoons. IRAS satellite measurements were used by
\citet{WC89a} to elucidate the properties of 75 Galactic sources.  While
the field UCH~II regions are typically excited by a single OB star, there
is evidence that other lower mass ``companions'' or small clusters are
present also \citet{Ket94}. The radio radiation is typically, but not
always, optically thick longward of 6 cm wavelengths (positive slope
$\alpha$ where the flux $S_{\nu} \propto~{\nu}^{\alpha}$).  Using the list
of Galactic UCH~II sources in Wood \& Churchwell that have their measured
radio wavelengths (e.g., 6 and 2~cm) we find that roughly 20\% are
optically thin (5 out of 24 sources), the rest are optically thick. 

Spherically symmetric models of UCH~II regions have been constructed by
\citet{Cet90}. Immediately surrounding the star is a small, low density
but very hot central ``cavity'' which underlays the denser ionized shell,
or H~II region, according to these models. This material is surrounded by
a thin, very dense neutral shock containing dust and gas.  Outside this
region is the large, warm dust cocoon, the source of the thermal emission,
and molecular hydrogen. The dust volume temperature is not single valued
as it is distributed over a large volume of space surrounding the star.
This material will radiate strongly in the sub-mm and mid-IR regions with
a maximum near $100~\mu$m from a mean T $\approx~30~^\circ$K. As most of
the dust cocoon is at temperatures less than $100~^\circ$K, only a minor
fraction of the overall thermal emission will be found shortwards of
$\approx~30~\mu$m.  At near IR wavelengths ($\approx~2~\mu$m) no thermal
emission would be expected from the dust cocoon. 
 
In addition to radiating at wavelengths characteristic of its temperature
distribution, the dust has a wavelength dependent opacity that affects the
emergent light of the underlying star. The UCH~II region dust opacity is
very high in the visible with a transition to the optically thin case
somewhere in near IR wavelengths (shorter than $\approx 7~\mu$m according
to the models of Churchwell et al. 1990). A survey of 63 radio selected
Galactic UCH~II regions was carried out by \citet{Het01} using $H-$band
photometry and $K-$band spectroscopy. Roughly 50\% of their objects showed
evidence of extended nebular Br$\gamma$ emission or had candidate hot
stellar sources visible in the $H-$band. \citet{Wet99}, using a smaller
sample of southern UCH~II regions, detected central stars in 66\% of them
in the near IR.  Using a sample of IRAS sources color selected to be
UCH~II regions ({\it i.e., candidates}), \citet{Ket00} detected OB stars
in one-third and ``emission-line'' stars in another one-third. The others
had only foreground stars in the field.  For actual UCH~II regions the
observations thus seem to suggest that {\it half} of the exciting objects
will be visible at $\approx~2~\mu$m. 

\subsection{Disc Geometry}

There may be a further complication affecting the observational properties
of UCH~II regions and their exciting stars if a disc is present (e.g.,
Garay \& Lizano 1999).  This material will likely be sufficiently close to
the star and its temperature hot enough to radiate in the $K-$band. This
radiation could swamp the stellar continuum, obliterating any stellar
spectroscopic signatures at this wavelength. This condition may initially
be detected with accurate $JHK$ photometry as hot stars with excess
$K-$band emission will appear to be too ``red'' in the classic $J-H$ {\it
vs.} $H-K$ color-color diagram. 

A residual disc geometry seems to be present for a number of very young
(YSO candidate) late O/early B type stars in M17 \citep{Het97}. These
stars have anomalous $J-H$ {\it vs.} $H-K$ colors. With the exception of
Br$\gamma$ emission, their $K-$band spectra are otherwise featureless (or
have CO emission lines). The interstellar extinction towards M17 is
sufficiently low that far-red and blue spectra were able to be obtained.
There the disc emission becomes less prominent and underlying stellar
absorption features are seen, although they appear ``veiled''.
Furthermore, some hydrogen emission lines are seen to be double peaked, a
nearly certain indication of a disc morphology. Infall or outflow from
these discs is not yet established, nor can such motion be ruled out. 
Most of the stars with discs could be classified as late O or early B
type. In this cluster, the earliest type O stars do not show any evidence
of discs. This could merely reflect a time scale issue: for example, the
disc dissipation could depend inversely on the mass such that the most
massive stars in M17 have already lost their discs. 

Anomalous $JHK$ colors and, in some cases ``featureless'' $K-$band spectra
with Br$\gamma$ emission were found for several stars we have studied in
the GH~II regions W43, W42 and W31 \citep{Bet99,Bet00,Bet01}. Given near
IR properties similar to those found for the stars of M17, these objects
are believed to harbor remnant stellar discs.  Several OB stars with
$K-$band photometric or spectroscopic evidence of discs appear to excite
UCH~II regions in W31. In at least two field Galactic UCH~II regions,
G29.96+0.6 (Watson \& Hanson 1997) and G23.96+0.15 (Hanson et al. 2001),
photospheric lines from the central exciting O stars are clearly visible
in the $K-$band and are classifiable.  In a few other cases, Hanson et al.
find exciting stars with featureless spectra, which again is consistent
with veiling by a hot circumstellar material as if they have stellar
discs.  Additional work in this area is clearly needed. 

Discs are visible in {\it some} hot luminous stars well after the cocoons
have been evaporated. It seems obvious that if discs are present after the
end of the cocoon phase, {\it they must have been present throughout it}.
This follows logically since the initial stages of massive star birth
necessarily involve a massive accretion disc. Evidence for disc morphology
in this precursor ``hot core'' phase has been given by \citet{Cet98}. It
is hard to imagine a reason for this disc to disappear for awhile and then
another one to reappear near the end of the birth processes. It is not
impossible that all massive stars have discs during their initial birth
processes and this morphology persists, at least for some of them, to
beyond the UCH~II phase. Most, if not all, might have discs during the
entire UCH~II lifetime. Models of a such ``photo-evaporative'' discs have
been promoted by \citet{Het94}. 

In M17 some late O/early B type stars have discs, but others do not. This
distinction cannot be related to the final stellar mass.  A highly
speculative notion is to imagine that the late retention of remnant discs
is somehow related to whether or not a star is binary or single.  Radial
velocity studies of the binary frequency of the OB stars in M17 are
currently underway by PSC and other of his colleagues to address this
issue. 

\section{RADIO AND NEAR IR IMAGING OF W49A}

\subsection{Radio Map of W49A}

\begin{figure}
\plotone{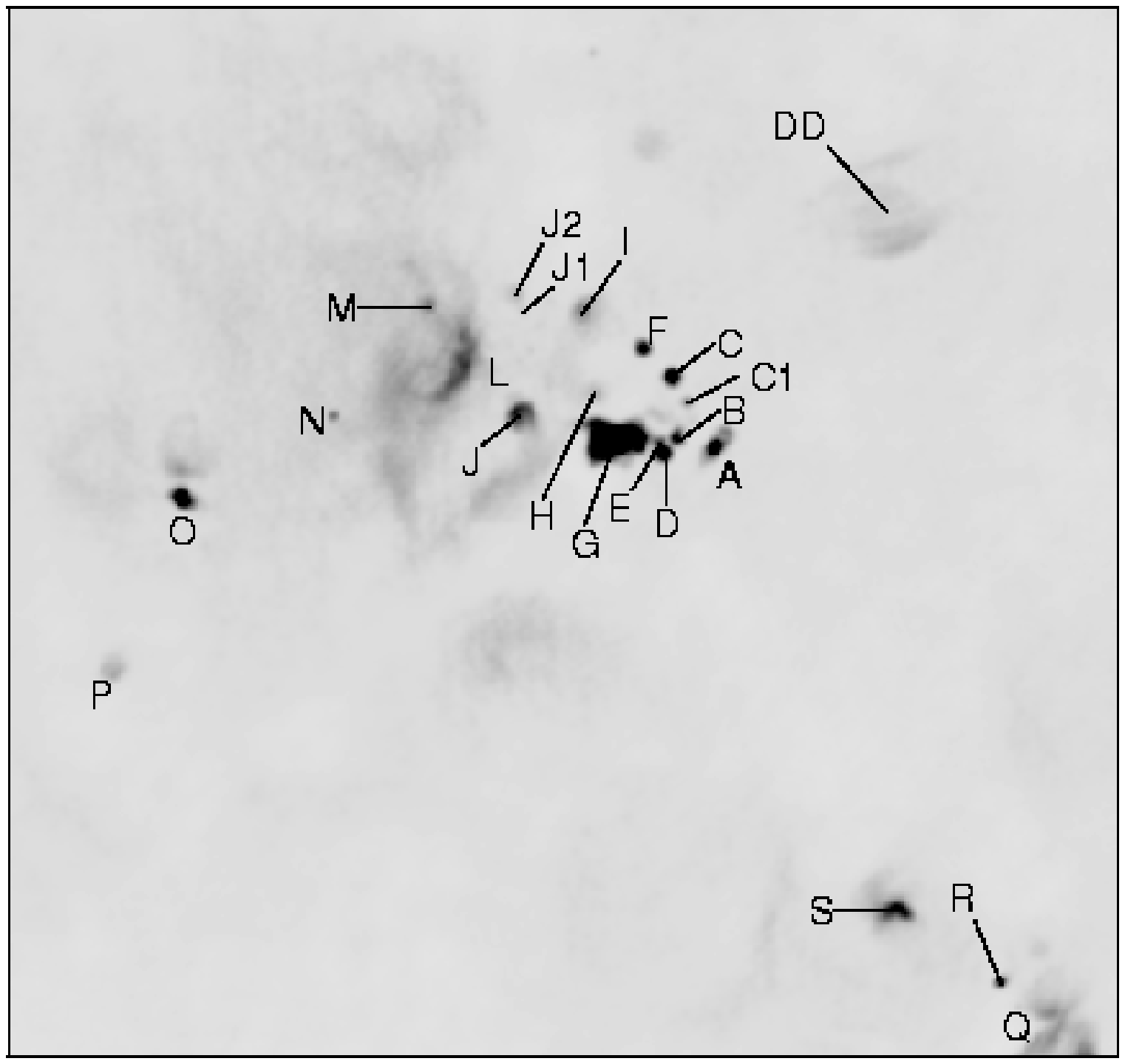}
\figurenum{1}

\figcaption[]{Radio continuum image of W49A kindly provided by C. DePree;
see \citet{Deet97}. North is up, East to the left. The image
is sampled at 0.20$''$ pix$^{-1}$ (beam size of 0.8$''$ FWHM) 
and the region shown is approximately 1.8$'$$\times$1.7$'$.
The radio sources from
\citet{Deet97} are marked. Source ``G'' contains multiple components.}

\label{radio}
\end{figure}

Figure~\ref{radio} shows a 3.6~cm radio grayscale map of the W49A
region, which we have adapted from the data presented by
\citet{Deet97}. 
The UCH~II point sources identified with letters by \citet{Det84} are
concentrated in the center of the W49A region; given their
number they might be thought of as a cluster of OB stars (Dreher et
al.). Towards the SW and NW of the central cluster are extended
(``resolved'') radio sources which have been labeled by
\cite{Deet97}. These are more properly thought of as H~II, or compact
H~II, regions. East of the cluster is amorphous and extended radio
emission, not all of which has been labeled by radio astronomers.

\subsection{K-band Image of W49A}

\begin{figure}
\plotone{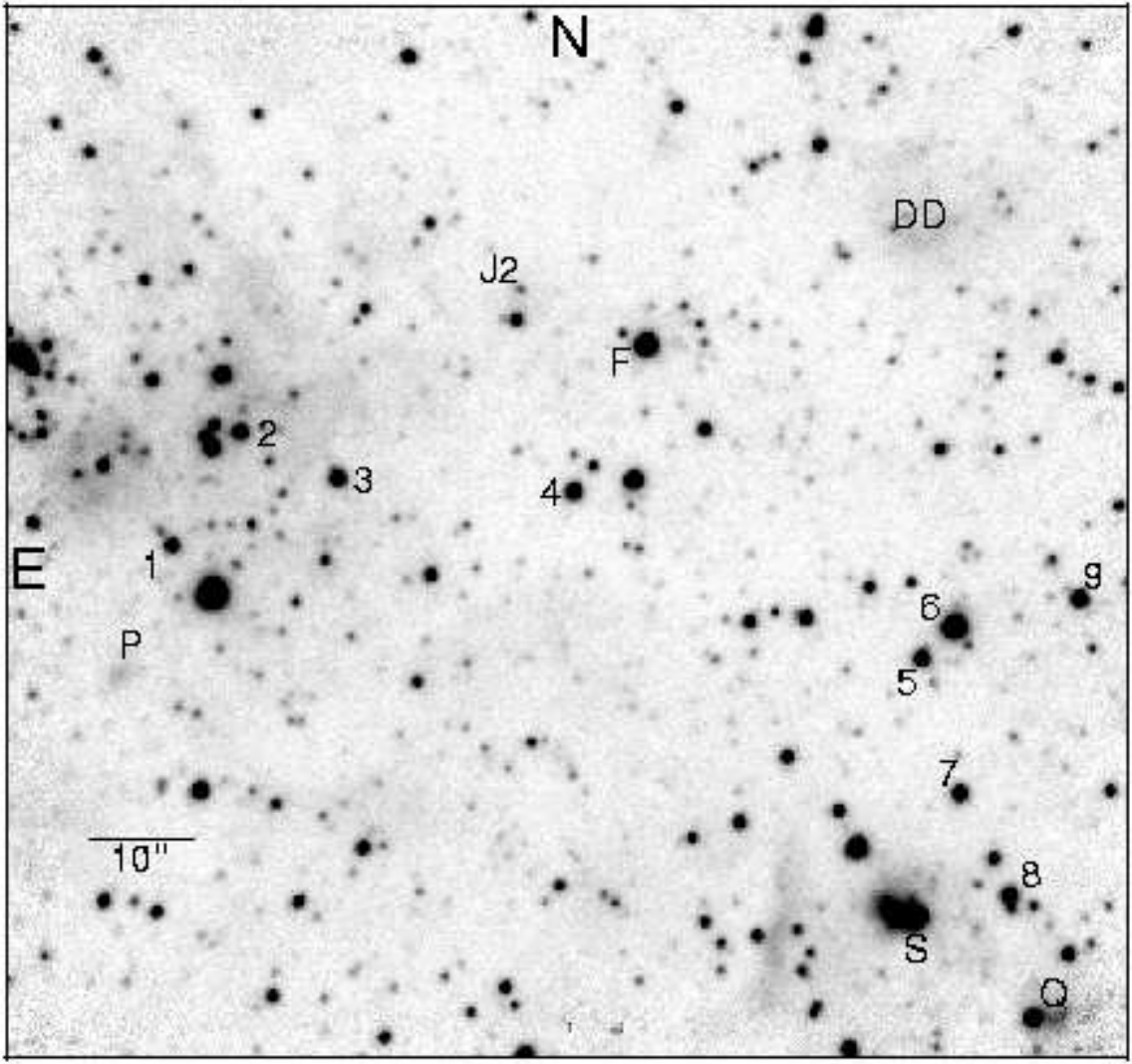}
\figurenum{2}

\figcaption[]{$K-$band image toward W49A. North is up, East to the
left. The OSIRIS image has a scale of 0.161$''$
pix$^{-1}$, and the $K-$band seeing was about 0.7$''$ FWHM.
The image is the same size as Figure~\ref{radio}.
Ultra-compact radio sources F and J2 are marked as are
the extended H~II regions P, Q, S, and DD; see
Figure~\ref{radio}. 
The remaining $K-$band objects (designated W49:CB01 1 to 9, and
labeled as ``1 to 9'') are candidate O and B
stars; see text.}

\label{kband}
\end{figure}

Using the Blanco 4-m telescope and facility IR imager, OSIRIS
\footnote{OSIRIS is a collaborative project between the Ohio State
University and CTIO. OSIRIS was developed through NSF grants AST 9016112
and AST 9218449.}, at Cerro Tololo Interamerican Observatory (CTIO), we
have obtained the $K-$band image of the W49A region shown in
Figure~\ref{kband}. The data taking and reduction are completely analogous
to that for the images described by \citet{Bet01}. The data were obtained
on the night of 20 May 2000 along with a similar $H-$band image.  The
total integration time was 225 seconds at $K$ and 350 seconds at $H$. The
data were obtained through thin cirrus.  A photometric calibration was
obtained on 11 July 2001 at the Blanco 4-m, again with OSIRIS. The short
exposures of W49 were calibrated using the standard star GSPC~S808-C 
(also known as [PMK98] 9178) from
\citet{peal98}. Photometry through 15 pixel radius apertures was used to
set the zero point of the 2001 images using four relatively bright and
uncrowded stars at $K$ and five at $H$. The same stars were then compared
to the 2000 data to flux calibrate the deeper images. The uncertainty in
the transformation for the deep images is taken as the standard deviation
in the mean of the four stars at $K$ and five at $H$ and was 0.027 $\%$
and 0.026 $\%$, respectively. 

The $K-$band filter includes contributions from nebular, presumably
predominantly Br$\gamma$, emission. There is no cluster visible in the
$K-$band image at the position of the UCH~II regions, nor is any
extended nebular contribution seen there.  Some combination of the dust of
the individual cocoons and that within the cluster is optically thick in
the $K-$band so the stellar light is totally absorbed. This cocoon
material must also be insufficiently hot to provide any thermal emission
at $2.2~\mu$m IR wavelengths. However, point sources F and J2 are clearly
detected in the $K-$band image (see also below). There is a connection
between several of the outlying H~II regions of Figure~\ref{radio} and
similar emission structure in Figure~\ref{kband}.  Sources Q and S to the
SW are clearly visible as bright extended objects, as is the fainter
object DD to the NW. Towards the East a number of stellar images seem to
be associated with the extended Br$\gamma$ and radio emission. It appears
that while the central stars of W49A are nearly all optically thick
to $K-$band radiation, in the periphery the extinction is much less and
the H~II regions and candidate exciting stars may be seen. 

\begin{figure}
\plotone{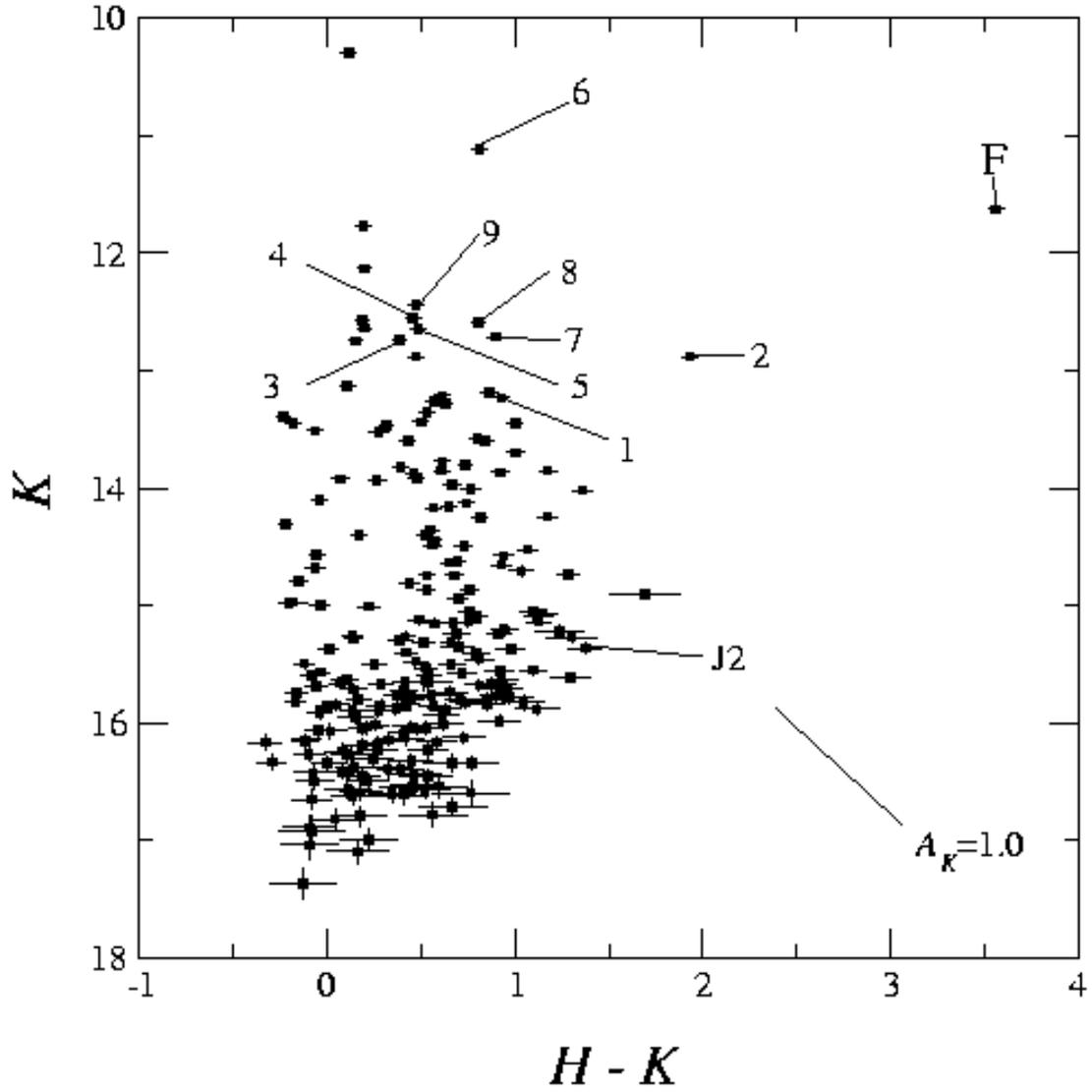}
\figurenum{3}

\figcaption[]{$K$ vs. $H-K$ color--magnitude diagram for the region of
Figure~\ref{kband}. The labeled sources W49:CB01 1 to 9 are candidate OB
stars; see text. Ultra--compact H~II radio sources F and J2 are also labeled.
The $K-$band uncertainties for the 
brighter stars are smaller than the plotted points.}

\label{cmd}
\end{figure}

The $K$ vs. $H-K$ color--magnitude diagram (CMD) is shown in
Figure~\ref{cmd}. The photometry was obtained using the
point--spread--function (PSF) fitting program DOPHOT \citep{sms93}. The
uncertainties are a combination of the PSF fitting error plus the
calibration uncertainty described above added in quadrature. There is
strong differential reddening evident in Figure~\ref{cmd}, but a
foreground sequence is seen at the bluest colors. A broad intermediate
sequence is seen at colors between about 0.5 and 1.0 magnitudes
corresponding to 1 \apge $A_K$ \apge 1.5 mag using the interstellar
extinction curve of \citet{m90}. A number of $K-$band sources 
fall in this range, and we believe they are candidate OB stars
that have sufficiently shed their natal material and may now be ionizing
the peripheral HII regions in W49A.
Following IAU convention, we designate these nine new sources 
as W49:CB01~1~to~9. Hereafter, (e.g. Figures~\ref{kband} and \ref{cmd})
we refer to them as sources ``1--9.''
If these are early O-type stars, their $A_K$ and $K-$band
magnitudes put them near to the canonical 11.4 kpc distance of W49A. 

\begin{figure}
\plotone{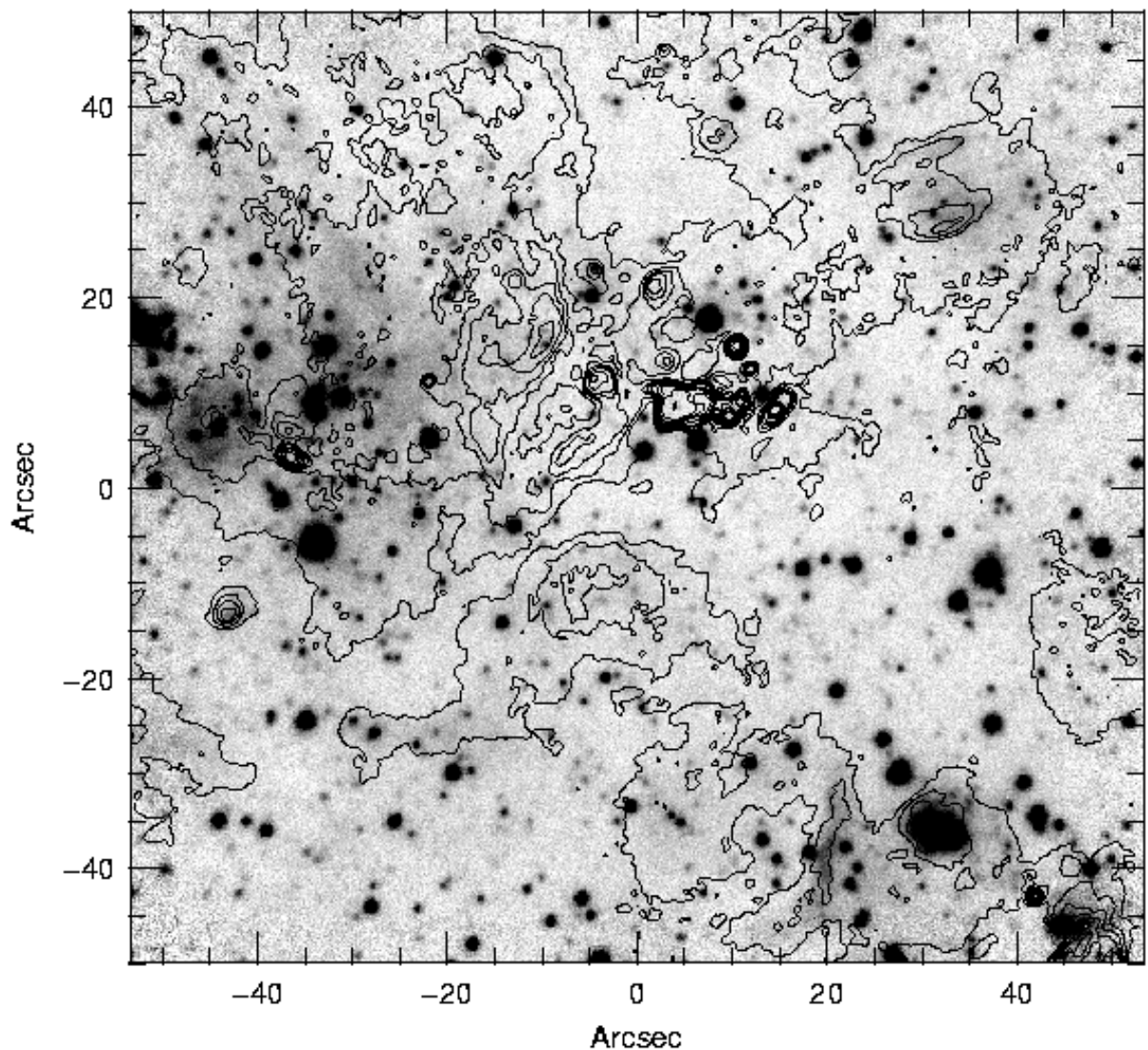}
\figurenum{4}

\figcaption[]{Overlay of $K-$band grayscale image (Figure~\ref{kband})
with the radio continuum image (Figure~\ref{radio}) contours (same size
as Figure~\ref{radio}). 
The center of the $K-$band image [0,0] is located at $\alpha$ (2000)~$=$
19$^{\rm h}10^{\rm m}13.30^{\rm s}$, $\delta$~(2000) $=$ 
09$^{\rm \circ}06^{'}03.5^{''}$.  Only
compact radio sources F and J2 from the central part of the
image have near infrared counter parts. Some of the $K-$band sources
at the periphery are candidate OB stars which have broken out from
their cocoons and now excite the more extended HII regions; see text
and Figure~\ref{kband}.}

\label{over}
\end{figure}

In Figure~\ref{over} we have plotted the radio contour overlay on the
$K-$band image. Point radio sources F and J2 are coincident with $K-$band
images. Object F is extremely red in $H-K$ and J2 is somewhat red compared
to the main body of stars with $H-K$ between 0.5 and 1. Source 2 is also
very red but it is not a radio source. The $K-$band magnitude of object F
might be anomalously bright if it has a reprocessing circumstellar disc. 
Following \citet{Het92}, the maximum $\Delta~H-K$ is 0.5 and $\Delta~K$ is
4.0 for an O star excited disc viewed face on (their Table 4). Then from
Figure~\ref{cmd} the stellar $H-K$ of F \apge 3 and $A_K$ \apge 4.8. 

A careful perusal of Figure~\ref{over} reveals that the rest of the UCH~II
``ring'' sources are not detected in the $K-$band. A conservative
estimate is that their $K-$band magnitudes must be fainter than 18. For the
canonical DM of 15.3 and adopting a ZAMS $M_K$ of -4.8 (type O3) for the
brightest stars we have $A_K$ \apge 7.5 for the central cluster. 

Our own unpublished spectra of sources F and J2 from a recent Kitt Peak
run (principal investigator Dr. Jeff Goldader), which will be presented in
a forthcoming article, suggest both stars are ``featureless'' in the
$K-$band, with the exception of nebular (or circumstellar) Br$\gamma$
emission. (Given that neither object has stellar CO absorption features we
can immediately conclude that they are not K or M stars in the
foreground.) The description of the spectra is similar to that found for
the brightest star (\#~1) in W31 \citep{Bet01}, which is also the exciting
star for an UCH~II region. Again we have the situation in which there is
evidence consistent with a disc and a dust cocoon that is optically thin
at $2.2~\mu$m. The diffuse radio source P (to the SE) is also extended, or
nebular, in the $K-$band image. We plan to obtain $K-$band spectra of the
candidate exciting stars with our colleagues in Brazil in the near future.
This should enable us to classify the exciting stars and potentially to
ascertain their ages and use them for an accurate spectrophotometric
distance determination to W49A as we have done previously in W43, W42, and
W31 \citep{Bet99,Bet00,Bet01}. 

\subsection{Thermal IR Imaging}

Relatively high spatial resolution 10--20~$\mu$m imaging of W49A has
recently been reported by \cite{Set00}. At these wavelengths one is
measuring the re--radiated stellar energy from the warm dust surrounding
the stars. Many of the radio continuum sources were detected, as would be
expected if there were luminous stars buried within their cocoons.
The thermal IR point sources detected included F, G (which is multiple),
I, J, J2, L, M, R and S; sources H, J1 and K were not detected
shortward of 20~$\mu$m. A number of extended sources (mostly with double
letters - \citet{Deet97}) were also detected at 10-20~$\mu$m. 

However, the western most sources, A, B, C, C1, D and E, were not detected
at 20~$\mu$m or shorter wavelengths.  \citet{Set00} interpret these stars
as lying behind a very opaque local dust cloud blocking our line of sight,
such that even the short wavelength mid IR emitted by the dust of the
individual UCH~II regions is absorbed.  Given their location together on
the west side of the ``ring'', this possibility seems reasonable. The
non-detection of these sources in the $K-$band is then predictable.
Following the \citet{Set00} estimate that the local cloud reduces the
20$\mu$m flux by a least a factor 5, then we have $A_{20} = -2.5log5 =
1.75, A_{20}/A_K= 0.196$, according to \citet{m90}, and $A_K$ \apge 8.9. This
estimated extinction for the western ``ring'' sources is somewhat larger
than our lower limit ($A_K$ \apge 7.5) derived above for the non-detection of
any of the ``ring'' (aside from F), which makes sense.

\section{INTERPRETATION OF THE EVOLUTIONARY STATE OF W49A}

For an individual UCH~II region the surrounding gas and the dust should be
evolving on relatively rapid time scales as the ionized gas volume is
expanding and the dust is being evaporated and dissipated by the exciting
star. Stellar winds will play a role too. \citet{WC89b} give statistical
arguments based upon their number counts of UCH~II regions and those of OB
stars in the Galaxy that suggest the ages of the former are 10 - 20 \% of
the main sequence lifetime. Thus the cocoon stages could last up to 2
million years.  It is well known that such lifetimes are much larger than
simple dynamical arguments would suggest but a consensus explanation for
this discrepancy is not yet available. In any case, given that there is an
UCH~II region, the underlying star must be hot and already in the main
sequence hydrogen-burning phase. 

Even after the buried star has begun its initial main sequence evolution
it may continue to retain a disc. It was argued above (\S2.2) that an
inner disc may be present {\it throughout} the cocoon phase; such a disc
is likely to be ``photo-evaporative'' \citep{Het94}. The
existence of this disc and the complications arising from its evolution
and eventual destruction via dissipation and/or accretion to the star
might well substantially lengthen the UCH~II phase (Welch 1993). A full 
radiative hydrodynamical treatment of the non-spherically symmetric and
inhomogeneous dust and gas surrounding a massive star at this phase of its
evolution is still well beyond current understanding.  In any case, with
time the densities of the gas and of the dust will rapidly diminish,
probably with different time scales. While the physical conditions of
these materials are related, their optical and geometrical depths may not
be closely coupled.  As the cocoon is being dissipated, its optically thin
transition point will begin to shift, and the star will be seen to
``emerge'' from its birth place at progressively shorter near IR
wavelengths.  Similarly, the inner ionized hydrogen region will expand,
becoming first optically thin and eventually transitioning to a
``compact'' and eventual ``classical'' H~II radio source. 

What do we observe in W49A?  Let us initially set aside the central
sources A, B, C, C1, D, and E, for which there is no mid--IR wavelength
thermal emission because of the high opacity of a localized dust cloud
(and no $K-$band emission either for the same reason).  For the remaining
UCH~II regions, there are then two subsets as characterized by their near
IR properties.  In the first of these, the warm dust cocoons emit thermal
IR radiation but they remain opaque at 2.2~$\mu$m. The accretion disc is
buried within the cocoon and not observed (even though it may be hot
enough to emit in the $K-$ band). In the central cluster of W49A, there
are eight examples of these UCH~II sources: five in G, I, J, \& M. In the
second subset of UCH~II regions, thermal {\it and} $K-$ band emission from
the central object is found. This is the situation for sources F \& J2, in
the central cluster. In F \& J2, the $K-$band spectrum
is featureless but with nebular (circumstellar?) emission lines. In these
stars, the cocoon has become transparent down to the central object whose
putative disc radiates in the $K-$band. The fraction in each phase is
consistent (given the very small numbers here) with the 50\% statistics of
$K-$band detection of the central objects from \citet{Het01} for the field
Galactic UCH~II regions. These sources are illuminating UCH~II
regions, so that while the stars are beginning to ``emerge'' from the
cocoons, their ionized hydrogen region remains tightly bounded. 

\citet{Deet00} have obtained multiwavelength VLA and BIMA observations
of several of the point sources within the central cluster.  It is
interesting that sources C, D, and F have ``flat'' spectral indices
($\alpha$) between 1.3 cm and 3.3 mm while A, B1, B2, G1, G2a and
G2b are ``rising''. These are the optically ``thin'' and ``thick''
cases for the free-free radio emission, respectively.  Given that F is
detected in the $K-$band, and its radio emission is also optically
thin, it is tempting to suggest that perhaps sources C and D would
also be visible in the $K-$band.  Unfortunately, they remain buried in
the intense local dust cloud on the west side of the ring.

Finally there are the hot star candidates in the periphery of the cluster
that are associated with H~II, rather than UCH~II, regions. The ionized
surroundings of these stars are no longer physically constrained by the
overlying dust cocoons, which have now dissipated. There may still be
remnant discs, as with some of the stars in M17, but the ultra compact
H~II regions are no more. $K-$band spectroscopy is now needed to ascertain
the spectral types and to determine the presence or absence of discs.  On
the basis of the lack of UCH~II regions here, as compared to the center,
we might suggest that star formation began in the periphery.
Alternatively, the lifetime of the UCH~II/cocoon stages might be shorter
in a less dense external interstellar environment. 

\section{A CONNECTION TO SIMILAR PHENOMENA IN OTHER GALAXIES}

\citet{Jet01} have discovered compact thermal radio sources in the nearby
galaxies M33, NGC~253, and NGC~6946 from an examination of papers in the
literature that surveyed these galaxies for SNR.  (SNR show non--thermal
radio continua, so they are easy to disentangle from other sources.) The
listed thermal sources, which are also optically thick in cm wavelengths,
emit a number of Lyman continuum photons equivalent to between a few and a
few hundred O stars. An examination of the corresponding optical fields
shows little or no visible radiation at the positions of half of the radio
sources; in most of the other half, the stellar contributions are
insufficient to account for the implied numbers of O type stars. These
UDH~II sources are scaled up in radio luminosity from the (mostly single
star excited) field UCH~II regions in our Galaxy. \citet{Jet01} suggest
these objects to be buried open clusters and associations. W49A can be
considered to be one of this class of object, a newly born counterpart of
luminous optically visible GH~II regions such as NGC~604 in M33.  These
buried objects should also be strong thermal IR emitters, as are the
UCH~II regions in W49A. Their visibility in the K-band will give clues about
the extent to which the cluster has begun to emerge from the natal dust.

\section{CONCLUSIONS}

Near IR imaging of the Galactic GH~II region W49A has been presented. Most
of the components of the previously known cluster of UCH~II regions are
invisible at these wavelengths. The cocoons of these stars are still
sufficiently opaque at near IR wavelengths to absorb radiation from the
central objects, and most of them are known to radiate in the thermal IR.
These may be the least evolved type of UCH~II region.

Two UCH~II point sources in the central cluster have $K-$band
counterparts;  unpublished $K-$band spectra show no stellar absorption
lines but Br$\gamma$ circumstellar(?) emission features are seen.
This spectrum is consistent with an origin in a hot, dense disc. Stars
with similar characteristics have previously been found to be exciting
UCH~II regions in other Galactic GH~II regions. The birth cocoons of these
objects have thinned sufficiently that the putative disc emission from the
central object is visible. We suggest that an accretion disc must be
present {\it throughout} the entire UCH~II region phase of massive star
birth if the revealed object has a remnant disc. 

There is evidence of nebular Br$\gamma$ emission and candidate exciting
stars around the periphery of the cluster. The outer parts of W49A might
well resemble a normal H~II region if viewed without the intervening
interstellar dust: recall the low density radio source descriptions 
of \citet{Met67} and of \citet{Det84}. It appears that star birth
may have begun on the edges of W49A and shifted more recently to the denser
interstellar medium in the center. Alternatively, the lifetime of the 
UCH~II/cocoon phases may lengthen in the denser environment of the 
central cluster. 

It has long been suggested that massive stars form nearly entirely in
clusters and associations. In W49A, we see an example of the earliest
parts of the birth processes, with the individual stars in various stages
of emergence from their cocoons. In W49A the newly born massive stars form
within their own cocoons and retain their individuality. As with field
UCH~II regions, small mass stars may be present close by.

W49A represents a Galactic GH~II region still going through its initial
birth pangs.  It is similar in luminosity ($\approx~100$ O star
equivalents) to the brightest UDH~II regions recently discovered in M33,
NGC253 and NCG6946 \citep{Jet01}. Being relatively nearby, it will serve
as a ``stepping stone'' towards understanding massive star birth in more
distant environments.

ACKNOWLEDGEMENTS

We are pleased to recognize the help received from Jeff Goldader
in obtaining $K-$band spectra of some objects within W49A. We thank
Margaret Hanson for a preprint of her important paper concerning
near infrared observations of Galactic UCH~II regions, and Chris DePree 
for a electronic version of his radio imaging data.  Useful comments 
on the initial draft of this manuscript have been received from  
Ed Churchwell, Kelsey Johnson, and Bill Vacca. PSC appreciates continuous 
support from the NSF.


\end{document}